\documentclass[reqno,11pt]{amsart}

\usepackage{epsf}

\def\be{\begin{equation}}
\def\ee{\end{equation}}
\def\ba{\begin{align}}
\def\bm{\begin{multline}}
\def\bfig{\begin{figure}[htb]}
\def\efig{\end{figure}}
\newcommand{\bibit}[1]{\bibitem[#1]{#1}}
\newcommand{\paper}[1]{{\it #1}, }
\newcommand{\journal}[4]{#1 #2, #3 (#4)}
\newcommand{\CMP}{Commun.\ Math.\ Phys.}
\newcommand{\HPA}{Helv.\ Phys.\ Acta}
\newcommand{\JSP}{J.\ Stat.\ Phys.}

\newtheorem{theorem}{Theorem}%[section]

\newcommand{\nn}{\nonumber}

\DeclareMathSymbol{\leqslant}{\mathalpha}{AMSa}{"36}
\DeclareMathSymbol{\geqslant}{\mathalpha}{AMSa}{"3E}
\DeclareMathSymbol{\doteqdot}{\mathalpha}{AMSa}{"2B}
\DeclareMathSymbol{\circlearrowright}{\mathalpha}{AMSa}{"08}
\DeclareMathSymbol{\subsetneq}{\mathalpha}{AMSb}{"28}
\DeclareMathSymbol{\supsetneq}{\mathalpha}{AMSb}{"29}
\renewcommand{\leq}{\;\leqslant\;}
\renewcommand{\geq}{\;\geqslant\;}

\newcommand{\dd}{{\rm d}}
\newcommand{\e}[1]{\,{\rm e}^{#1}\,}

\newcommand{\sumtwo}[2]{\sum_{\substack{#1 \\ #2}}}

\def\Tr{{\operatorname{Tr\,}}}

\newcommand{\const}{{\text{\rm const}}}
\newcommand{\upchi}{\raise 2pt \hbox{$\chi$}}
\def\writefig#1 #2 #3 {\rlap{\kern #1 truecm \raise #2 truecm
\hbox{#3}}}
\def\figtext#1{\smash{\hbox{#1}} \vspace{-5mm}}
\newcommand{\caA}{{\mathcal A}}

\newcommand{\caC}{{\mathcal C}}

\newcommand{\caF}{{\mathcal F}}
\newcommand{\caG}{{\mathcal G}}

\newcommand{\bbA}{{\mathbb A}}

\newcommand{\bbR}{{\mathbb R}}

\newcommand{\bbZ}{{\mathbb Z}}

\newcommand{\bsomega}{{\boldsymbol \omega}}

\begin{document}

{\hfill\small Moscow Mathematical Journal {\bf 4} 511--522 (2004)}
\vspace{2mm}

\title{Cluster expansions \& correlation functions}

\author{Daniel Ueltschi}

\address{Daniel Ueltschi \hfill\newline
Department of Mathematics \hfill\newline
University of Arizona \hfill\newline
Tucson, AZ 85721, USA\hfill\newline
{\small\rm\indent http://math.arizona.edu/$\sim$ueltschi}}
\email{ueltschi@math.arizona.edu}

\maketitle

\vspace{-5mm}

\begin{centering}
{\small\it
Department of Mathematics, University of California, Davis\\
}
\end{centering}

\vspace{2mm}

\begin{quote}
{\small
{\bf Abstract.}
A cluster expansion is proposed, that applies to both continuous and discrete systems. The
assumption for its convergence involves an extension of the neat Koteck\'y-Preiss
criterion.
Expressions and estimates for correlation functions are also presented. The results are applied to
systems of interacting classical and quantum particles, and to a lattice polymer model.

\vspace{1mm}

}  % end \small

\vspace{1mm}

\vspace{1mm}
\noindent
{\footnotesize {\it 2000 Math.\ Subj.\ Class.:} 82B05, 82B10.}
\end{quote}

\vspace{1mm}

\section{Introduction}

Cluster expansions were introduced at the dawn of Statistical Mechanics for the study of high
temperature gases of interacting particles. They constitute a powerful perturbative method that is
suitable for geometrically large systems, such as encountered in Statistical Physics. Numerous
articles have contributed to the subject; a list of relevant publications is
\cite{GK,Cam,Bry,KP,Pfi,Dob,BZ,Mir,Far} and
references therein. Cluster expansions can now be found in standard books, see Chapter 4 of Ruelle
\cite{Rue}, or Chapter V of Simon \cite{Sim}. They apply to continuous systems such as classical or
quantum models of interacting particles, and also to discrete systems such as polymer models, spin
models (with discrete or continuous spin spaces), or lattice particle models. The methods
for treating these various situations share many similarities, but a somewhat different cluster
expansion was so far required in each case.

The exposition of the cluster expansion is often intricate, and the paper of Koteck\'y and
Preiss \cite{KP} should be singled out for proposing a clear and concise theorem, that involves a
neat criterion for the convergence of the expansion. This theorem applies to discrete systems only,
and its rather difficult proof was subsequently simplified in \cite{Dob,BZ,Mir}.
Explicit expressions for the contribution of cluster terms can be found in the article of Pfister
\cite{Pfi}; this helps
clarifying the situation and allows for computations of lowest order terms. The condition for the
convergence is the same as in \cite{KP} in the case where polymers are subsets of a lattice.

The goal of this paper is to present a general theorem that applies to both continuous and discrete
systems. The condition for the convergence is given by an extension of the criterion of Koteck\'y
and Preiss, see Equation \eqref{KPcrit1} below, and the contribution of cluster terms involves explicit
expressions. Furthermore, we derive expressions and
estimates for correlation functions.

The theorems presented here are illustrated in Section \ref{secill} in three different situations,
namely classical and quantum gases of interacting particles, and lattice polymer models.

\section{Cluster expansions}

Let $(\bbA, \caA, \mu)$ be a measure space; $\mu$ is a complex measure and 
$|\mu|(\bbA)<\infty$, where $|\mu|$ is the total variation (absolute value) of $\mu$.
Let $\zeta$ be a complex measurable symmetric function on $\bbA\times\bbA$.
The {\it partition function} $Z$ is defined by
\be
\label{deffpart}
Z = \sum_{n\geq0} \frac1{n!} \int\dd\mu(A_1) \dots \int\dd\mu(A_n) \prod_{1\leq
i<j\leq n} \bigl( 1 + \zeta(A_i,A_j) \bigr).
\ee
The term $n=0$ of the sum is understood to be 1.

We denote by $\caG_n$ the set of all (unoriented)
graphs with $n$ vertices, and $\caC_n \subset \caG_n$ the set of connected graphs of $n$ vertices.
We introduce the following combinatorial function on finite sequences $(A_1,\dots,A_n)$ of $\bbA$:
\be
\varphi(A_1,\dots,A_n) = \begin{cases} 1 & \text{if } n=1 \\ \frac1{n!} \sum_{G \in \caC_n} \prod_{(i,j)
\in G} \zeta(A_i,A_j) & \text{if } n\geq2. \end{cases}
\ee
The product is over edges of $G$. A sequence
$(A_1,\dots,A_n)$ is a {\it cluster} if the graph with $n$ vertices and an edge between $i$ and $j$ whenever
$\zeta(A_i,A_j) \neq 0$, is connected.

The cluster expansion allows to express the logarithm of the partition function as a sum (or an
integral) over clusters.

\begin{theorem}[Cluster expansion]\hfill
\label{thmclexp}

\noindent
Assume that $|1+\zeta(A,A')|\leq1$ for all $A,A'\in\bbA$, and that there exists a nonnegative
function $a$ on $\bbA$ such that for all $A\in\bbA$,
\be
\label{KPcrit1}
\int\dd|\mu|(A') \, |\zeta(A,A')| \, \e{a(A')} \leq a(A),
\ee
and $\int\dd|\mu|(A) \e{a(A)} < \infty$. Then we have
$$
Z = \exp\Bigl\{ \sum_{n\geq1} \int\dd\mu(A_1) \dots \int\dd\mu(A_n) \, \varphi(A_1,\dots,A_n) \Bigr\}.
$$
Combined sum and integrals converge absolutely. Furthermore, we have for all $A_1\in\bbA$
\be
\label{b1}
1 + \sum_{n\geq2} n \int\dd|\mu|(A_2) \dots \int\dd|\mu|(A_n) \, |\varphi(A_1,\dots,A_n)| \leq
\e{a(A_1)}.
\ee
\end{theorem}

The rest of the section is devoted to the proof of this theorem; the reader interested in results only should jump
to Section \ref{seccorrfcts} that discusses correlation functions.

\begin{proof}
Another inequality turns out to be helpful. Multiplying both sides of \eqref{b1} by $|\zeta(A,A_1)|$
and integrating over $A_1$, we find using \eqref{KPcrit1}
\be
\label{b2}
\sum_{n\geq1} \int\dd|\mu|(A_1) \dots \int\dd|\mu|(A_n) \Bigl( \sum_{i=1}^n |\zeta(A,A_i)|
\Bigr) |\varphi(A_1,\dots,A_n)| \leq a(A)
\ee
for all $A\in\bbA$.

The strategy is to show inductively that \eqref{KPcrit1} implies \eqref{b1}.
Convergence of the cluster expansion follows and allows to prove Theorem \ref{thmclexp}.

We prove that the following holds for all $N$,
\be
1 + \sum_{n=2}^N n \int\dd|\mu|(A_2) \dots \int\dd|\mu|(A_n) |\varphi(A_1,\dots,A_n)| \leq
\e{a(A_1)}.
\ee
The case $N=1$ is clear and we consider now any $N$. The left side is equal to
\be
\label{uneborne}
1 + \sum_{n=2}^N \int\dd|\mu|(A_2) \dots \int\dd|\mu|(A_n) \frac1{(n-1)!} \Bigl|
\sum_{G \in \caC_n} \prod_{(i,j)\in G} \zeta(A_i,A_j) \Bigr|.
\ee
Let us focus on the sum over connected graphs $G$.
Removing all edges of $G$ with one endpoint on 1 yields a possibly disconnected graph $G'$. Let $(G_1,
\dots, G_k)$ be a sequence of connected graphs where $G_i$ has set of vertices $V_i$, $V_1 \cup \dots
\cup V_k = \{2,\dots,n\}$, and $V_i \cap V_j = \emptyset$ if $i \neq j$. Each sequence determines a graph
$G'$, and to each $G'$ corresponds $k!$ such sequences. See Fig.\ \ref{figcluster} for an illustration. Therefore
\bfig
\epsfxsize=70mm
\centerline{\epsffile{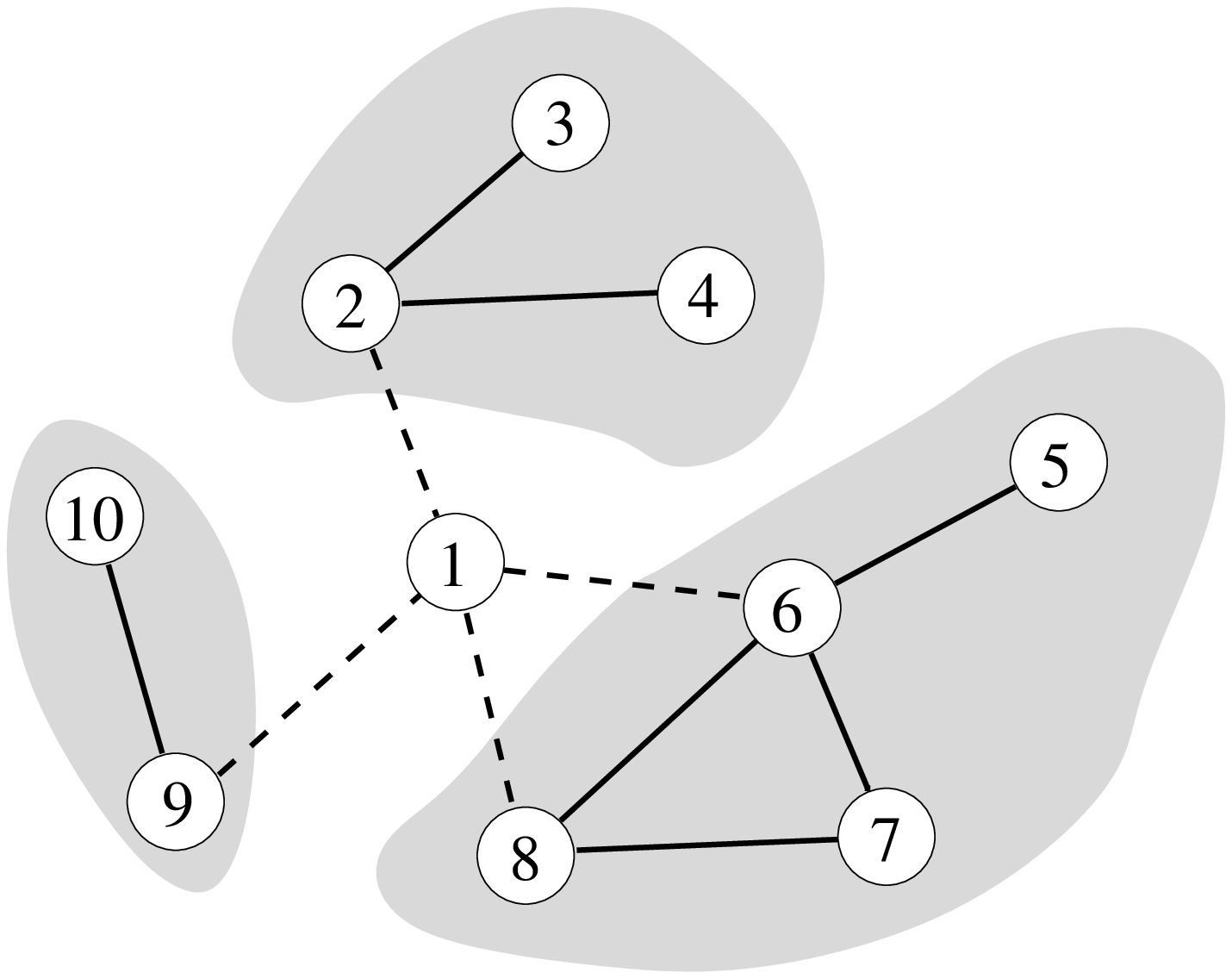}}
\figtext{
\writefig	8.2	5.5	{$G_1$}
\writefig	10.5	2.0	{$G_2$}
\writefig	3.8	1.6	{$G_3$}
}
\caption{Illustration for $G$, $G'$, and $(G_1,\dots,G_k)$.}
\label{figcluster}
\efig
\be
\Bigl| \sum_{G \in \caC_n} \prod_{(i,j)\in G} \zeta(A_i,A_j) \Bigr| \leq
\sum_{k\geq1} \frac1{k!} \Bigl| \sum_{(G_1,\dots,G_k)} \prod_{\ell=1}^k \Bigl\{ \prod_{(i,j)\in
G_\ell} \zeta(A_i,A_j) \sum_{G_\ell'} \prod_{(i,j)\in G_\ell'} \zeta(A_i,A_j) \Bigr\} \Bigr|.
\ee
The sum over $G_\ell'$ runs over nonempty sets of edges with one endpoint on
1 and one endpoint in $V_\ell$ ($i=1$ in the last product). We have
\be
\label{combi}
\sum_{G_\ell'} \prod_{(i,j)\in G_\ell'} \zeta(A_i,A_j) = \prod_{i\in V_\ell} \bigl( 1 +
\zeta(A_1,A_i) \bigr) - 1.
\ee
Using
\be
\label{combi2}
\prod_{i=1}^n (1+\alpha_i) - 1 = \Bigl[ \prod_{i=1}^{n-1} (1+\alpha_i) - 1 \Bigr] (1+\alpha_n) +
\alpha_n,
\ee
one easily proves inductively that the absolute value of \eqref{combi} is smaller than $\sum_{i\in V_\ell}
|\zeta(A_1,A_i)|$.

The sum over sequences $(G_1, \dots, G_k)$ can be done by first choosing the respective
numbers of vertices $m_1, \dots, m_k$ whose sum is $n-1$, then by summing over partitions of
$\{2,\dots,n\}$ in sets $V_1,\dots,V_k$ with $|V_i|=m_i$, and
finally by choosing connected graphs for each set of vertices. The number of partitions is
$\frac{(n-1)!}{m_1! \dots m_k!}$. Then \eqref{uneborne} can be bounded by
\bm
1 + \sum_{n=2}^N \sum_{k\geq1} \frac1{k!} \sumtwo{m_1,\dots,m_k\geq1}{m_1+\dots+m_k=n-1}
\prod_{\ell=1}^k \Bigl[ \int\dd|\mu|(A_1') \dots \int\dd|\mu|(A_{m_\ell}') \\
|\varphi(A_1',\dots,A_{m_\ell}')| \sum_{i=1}^{m_\ell} |\zeta(A_1,A_i')| \Bigr]. \nn
\end{multline}
We can sum over $n$; the constraint $m_1+\dots+m_k \leq N-1$
can be relaxed into $m_\ell \leq N-1$ for all $\ell$. Using \eqref{b2} with $n \leq N-1$,
we obtain the bound
\be
1 + \sum_{k\geq1} \frac1{k!} [a(A_1)]^k = \e{a(A_1)}.
\ee
This proves inequality \eqref{b1}. Absolute convergence of the cluster expansion follows from
\eqref{b1} and summability of $\e{a(A)}$.

The rest of the proof is standard and consists in showing that clusters are
indeed the terms of the expansion of the logarithm of the partition function. The idea is to expand
$Z$ so as to recognize an exponential.
\be
Z = 1 + \sum_{n\geq1} \frac1{n!} \int\dd\mu(A_1) \dots \int\dd\mu(A_n) \sum_{G
\in \caG_n} \prod_{(i,j)\in G} \zeta(A_i,A_j).
\ee
The sum over $n$ converges absolutely.
We proceed as above and consider sequences of connected graphs $(G_1,\dots,G_k)$ whose sets of
vertices form a partition of $\{1,\dots,n\}$. Summing first over the number of vertices of each
partition, then over partitions, we get
\ba
Z &= 1 + \sum_{n\geq1} \sum_{k\geq1} \frac1{k!}
\sumtwo{m_1,\dots,m_k\geq1}{m_1+\dots+m_k=n} \frac1{m_1! \dots m_k!} \nn\\
&\hspace{35mm} \prod_{\ell=1}^k \Bigl\{
\int\dd\mu(A_1) \dots \int\dd\mu(A_{m_\ell}) \sum_{G \in \caC_{m_\ell}} \prod_{(i,j)\in G} \zeta(A_i,A_j) \Bigr\} \\
&= 1 + \sum_{n\geq1} \sum_{k\geq1} \frac1{k!} \sumtwo{m_1,\dots,m_k\geq1}{m_1+\dots+m_k=n}
\prod_{\ell=1}^k \int\dd\mu(A_1) \dots \int\dd\mu(A_{m_\ell}) \, \varphi(A_1,\dots,A_{m_\ell}). \nn
\end{align}
Absolute convergence of the clusters allows to remove the sum over $n$, and
this completes the proof.
\end{proof}

\section{Correlation functions}
\label{seccorrfcts}

An advantage of the cluster expansion is to characterize correlation functions. The relevant
general expressions are
\ba
\label{corr}
Z(A_1,\dots,A_m) &= \sum_{n\geq m} \frac1{(n-m)!} \int\dd\mu(A_{m+1}) \dots \int\dd\mu(A_n)
\prod_{1\leq i<j\leq n} \bigl( 1 + \zeta(A_i,A_j) \bigr); \\
\hat Z(A_1,\dots,A_m) &= \sum_{n\geq m} \frac{n!}{(n-m)!}
\int\dd\mu(A_{m+1}) \dots \int\dd\mu(A_n) \, \varphi(A_1,\dots,A_n).
\label{defcorr}
\end{align}
It is understood that in both expressions, the case $n=m$ corresponds to taking the integrand
without integrating on polymers.

\begin{theorem}[Correlation functions]\hfill
\label{thmcorr}\nopagebreak

\noindent
Under the same assumptions as in Theorem \ref{thmclexp}, we have
$$
\frac{Z(A_1,\dots,A_m)}Z = \sum_{\{V_1,\dots,V_k\}} \prod_{j=1}^k \hat Z \bigl( (A_i)_{i \in V_j}
\bigr)
$$
where the sum is over partitions of $\{1,\dots,m\}$, i.e.\ $V_1 \cup \dots \cup V_k = \{1,\dots,m\}$,
and $V_i \cap V_j = \emptyset$ if $i\neq j$.
\end{theorem}

The next result deals with estimates of correlations. To exhibit a suitable decay, an efficient
strategy is to establish the criterion \eqref{KPcrit1} in a stronger form. We consider a nonnegative
function $b$ on $\bbA$, and a nonnegative symmetric function $c$ on $\bbA \times \bbA$ (both can be
identically zero, but the larger they are the better). We introduce the notation
\ba
&\mu_b(A) = \mu(A) \e{b(A)} \nn\\
&\zeta_c(A,A') = \zeta(A,A') \e{c(A,A')} \nn\\
& c(A_1,\dots,A_n) = \min_{G \in \caC_n} \sum_{(i,j) \in G} c(A_i,A_j) \quad \text{if } n\geq3 \nn\\
& \varphi_c(A_1,\dots,A_n) = \varphi(A_1,\dots,A_n) \e{c(A_1,\dots,A_n)}. \nn
\end{align}
The utility of functions $b$ and $c$ will be illustrated in Section \ref{secill}. The following
theorem contains estimates on correlations; compare with the definition \eqref{defcorr} of $\hat
Z(A_1,\dots,A_m)$.

\begin{theorem}[Decay of correlations]\hfill
\label{thmdecay}

\noindent
Assume that $|1 + \zeta_c(A,A')| \leq 1$ for all $A,A'\in\bbA$, and that there exists a nonnegative
function $a$ on $\bbA$ such that
\be
\label{KPcrit2}
\int\dd|\mu_b|(A') \, |\zeta_c(A,A')| \, \e{a(A')} \leq a(A)
\ee
for all $A\in\bbA$. Then the following estimate holds true for all $m\geq1$, and all $A_1,\dots,A_m
\in \bbA$,
\ba
\sum_{n\geq m} \frac{n!}{(n-m)!} &\int\dd|\mu_b|(A_{m+1}) \dots \int\dd|\mu_b|(A_n) \,
|\varphi_c(A_1,\dots,A_n)| \nn\\
&\leq \exp\Bigl\{ \tfrac1{m\gamma} \bigl[ (1+\gamma)^m - 1 \bigr] \sum_{i=1}^m
a(A_i) \Bigr\} \prod_{1\leq i<j\leq m} \bigl( 1 + |\zeta_c(A_i,A_j)| \bigr). \nn
\end{align}
Here, we set
\be
\label{defgamma}
\gamma = \sup_n \sup_{A_0,\dots,A_n} \Bigl| \prod_{i=1}^n \bigl( 1 + \zeta(A_0,A_i) \bigr) - 1 \Bigr|
\ee
(clearly, $0\leq\gamma\leq2$).
\end{theorem}

Notice that the case $m=1$ is
\be
\label{b3}
1 + \sum_{n\geq2} n \int\dd|\mu_b|(A_2) \dots \int\dd|\mu_b|(A_n) \, |\varphi_c(A_1,\dots,A_n)|
\leq \e{a(A_1)}
\ee
which is reminiscent of \eqref{b1}. Multiplying both sides of \eqref{b3} by $|\zeta_c(A,A_1)|$ and
integrating over $A_1$, we obtain from \eqref{KPcrit2}
\be
\label{b4}
\sum_{n\geq1} \int\dd|\mu_b|(A_1) \dots \int\dd|\mu_b|(A_n) \Bigl( \sum_{i=1}^n |\zeta_c(A,A_i)|
\Bigr) |\varphi_c(A_1,\dots,A_n)| \leq a(A)
\ee
for all $A\in\bbA$. \eqref{b3} and \eqref{b4} can be proved exactly the same way as \eqref{b1} and
\eqref{b2}.

We turn to the proofs of Theorems \ref{thmcorr} and \ref{thmdecay}. Readers interested in applications
should jump to Section \ref{secill}. We start with the proof of Theorem \ref{thmdecay} as it will
imply the convergence of the terms appearing in Theorem \ref{thmcorr}. We assume that \eqref{b4} has
been established.

\begin{proof}[Proof of Theorem \ref{thmdecay}]
A connected graph $G$ on $\{1,\dots,n\}$ can be decomposed into a graph $G'$ on $\{1,\dots,m\}$; a
partition $\{W_1,\dots,W_k\}$ of $\{m+1,\dots,n\}$; connected graphs $G_1,\dots,G_k$ where $G_i$
has set of vertices $W_i$; non-empty subsets $V_1,\dots,V_k$ of $\{1,\dots,m\}$; non-empty sets of edges
between $W_j$ and each vertex of $V_j$. This is illustrated in Fig.\ \ref{figcluster2}. Conversely,
choosing these graphs and sets of vertices yields a graph on $\{1,\dots,n\}$; it is
not necessarily connected, but we obtain an upper bound by dropping this
constraint. Therefore we get
\bfig
\epsfxsize=70mm
\centerline{\epsffile{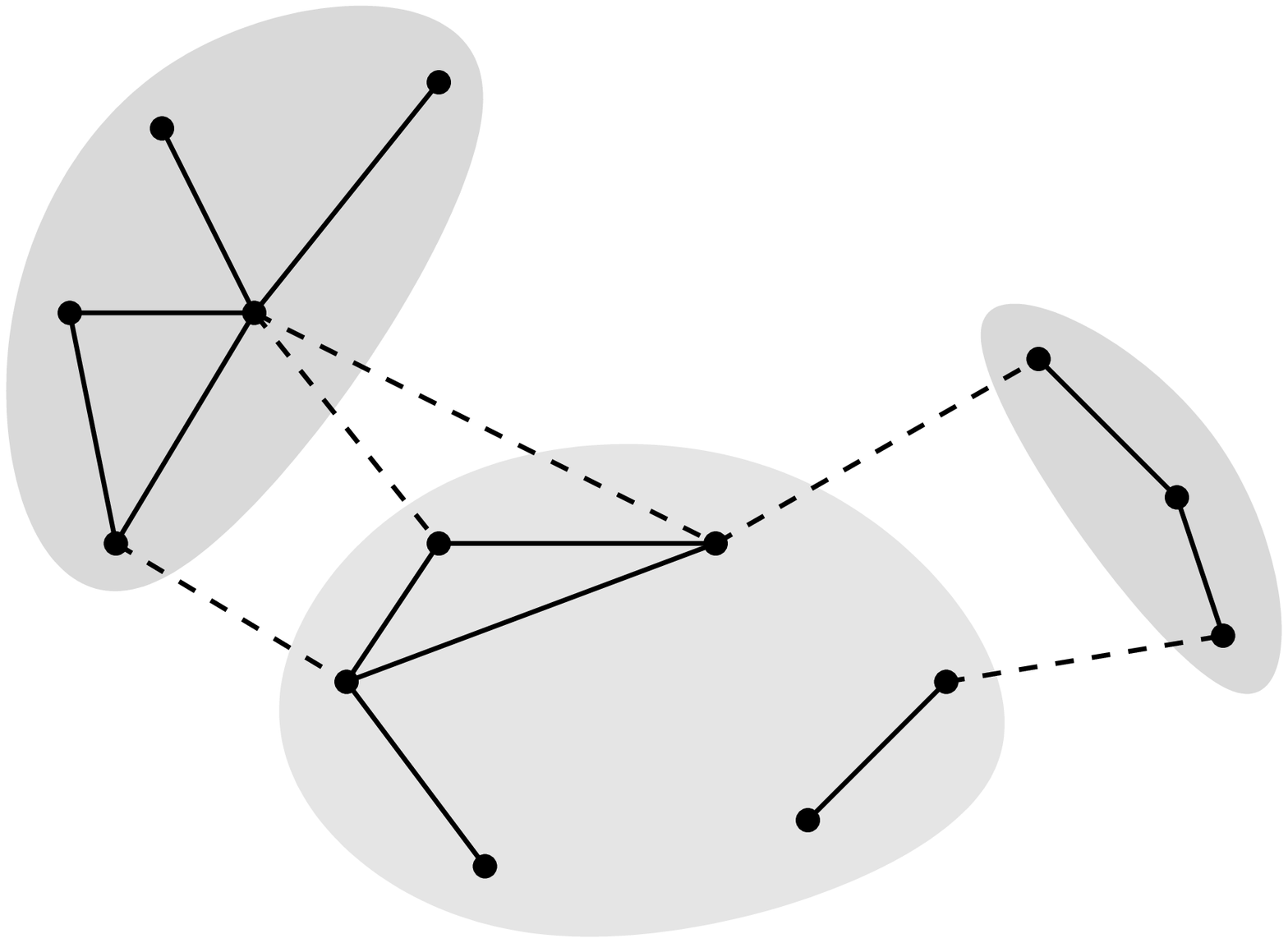}}
\figtext{
\writefig	3.4	3.5	{$G_1$}
\writefig	3.4	2.9	{$W_1$}
\writefig	10.7	2.9	{$G_2$}
\writefig	10.8	2.3	{$W_2$}
\writefig	7.2	1.9	{$G'$}
\writefig	6.4	1.4	{$\{1,\dots,m\}$}
}
\caption{}
\label{figcluster2}
\efig
\bm
\Bigl| \sum_{G\in\caC_n} \prod_{(i,j)\in G} \zeta(A_i,A_j) \Bigr| \e{c(A_1,\dots,A_n)} \leq
\sum_{G' \in \caG_m} \prod_{(i,j)\in G'} |\zeta_c(A_i,A_j)| \sum_{\{W_1,\dots,W_k\}} \\
\prod_{\ell=1}^k \biggl\{ \Bigl| \sum_{G_\ell \in \caC(W_\ell)} \prod_{(i,j)\in G_\ell} \zeta(A_i,A_j)
\Bigr| \e{c((A_i)_{i\in W_\ell})} \\
\sumtwo{V_\ell \subset \{1,\dots,m\}}{V_\ell \neq \emptyset} \exp\bigl\{ \min_{i\in V_\ell, j\in
W_\ell} c(A_i,A_j) \bigr\} \prod_{i \in V_\ell} \Bigl| \prod_{j\in W_\ell} \bigl( 1 + \zeta(A_i,A_j)
\bigr) - 1 \Bigr| \biggr\}.
\end{multline}

From the definition \eqref{defgamma} of $\gamma$, we have
\ba
\prod_{i\in V_\ell} \Bigl| \prod_{j\in W_\ell} \bigl( 1 + \zeta(A_i,A_j) \bigr) - 1 \Bigr| &\leq
\frac{\gamma^{|V_\ell|-1}}{|V_\ell|} \sum_{i\in V_\ell} \Bigl| \prod_{j\in W_\ell} \bigl( 1 +
\zeta(A_i,A_j) \bigr) - 1 \Bigr| \nn\\
&\leq \frac{\gamma^{|V_\ell|-1}}{|V_\ell|} \sum_{i\in V_\ell, j\in W_\ell} |\zeta(A_i,A_j)|.
\end{align}
We used \eqref{combi} and \eqref{combi2} for the second inequality. Furthermore
\be
\sumtwo{V_\ell \subset \{1,\dots,m\}}{V_\ell \neq \emptyset} \frac{\gamma^{|V_\ell|-1}}{|V_\ell|}
\sum_{i\in V_\ell, j\in W_\ell} |\zeta(A_i,A_j)| = \frac1{m\gamma} \bigl[ (1+\gamma)^m - 1 \bigr]
\sum_{i=1}^m \sum_{j\in W_\ell} |\zeta(A_i,A_j)|.
\ee
Then
\bm
\sum_{n\geq m} \frac1{(n-m)!} \int\dd|\mu_b|(A_{m+1}) \dots \int\dd|\mu_b|(A_n) \Bigl| \sum_{G\in\caC_n}
\prod_{(i,j)\in G} \zeta(A_i,A_j) \Bigr| \e{c(A_1,\dots,A_n)} \\
\leq \sum_{G'\in\caG_m} \prod_{(i,j)\in G'}
|\zeta_c(A_i,A_j)| \sum_{n\geq m} \sum_{k\geq0} \frac1{k!}
\sumtwo{n_1,\dots,n_k\geq1}{n_1+\dots+n_k=n-m} \\
\prod_{\ell=1}^k \biggl\{ \frac1{n_\ell!} \int\dd|\mu_b|(A_1') \dots
\int\dd|\mu_b|(A_{n_\ell}') \Bigl| \sum_{G_\ell \in \caC_{n_\ell}} \prod_{(i,j)\in G_\ell}
\zeta(A_i,A_j) \Bigr| \e{c(A_1',\dots,A_{n_\ell}')} \\
\frac1{m\gamma} \bigl[ (1+\gamma)^m - 1 \bigr] \sum_{i=1}^m \sum_{j=1}^{n_\ell} |\zeta_c(A_i,A_j)|
\biggr\}.
\end{multline}
The case $k=0$ of the right side should be understood as the case $n=m$ of the left side. The expression
inside brackets in
the right side converges because of \eqref{b4}; this allows to sum over $n$, and we get Theorem
\ref{thmdecay}.
\end{proof}

\begin{proof}[Proof of Theorem \ref{thmcorr}]
We expand the product in \eqref{corr} and we obtain a sum over graphs of $\caG_n$. A graph of
$\caG_n$ can be written as a sequence of graphs $(G_1,\dots,G_k,G')$, where each $G_i$ is connected
and has at least one vertex in $\{1,\dots,m\}$, and the vertices of $G'$ are in $\{m+1,\dots,n\}$.

Let $W_i$ be the set of vertices of $G_i$, $V_i = W_i \cap \{1,\dots,m\}$, and $W'$ be the set of
vertices of $G'$. $\{V_1,\dots,V_k\}$ is a partition of $\{1,\dots,m\}$ and $W' \subset
\{m+1,\dots,n\}$. Furthermore, let $n_j = |W_j \setminus V_j|$, $1\leq j\leq k$, and $p = |W'|$.

The number of partitions $\{W_1,\dots,W_k,W'\}$ of $\{1,\dots,n\}$ corresponding to given
$n_1$, \dots, $n_k$, $p$, and $\{V_1,\dots,V_k\}$, is equal to $\frac{(n-m)!}{p! \prod_i n_i!}$.
Therefore
\bm
Z(A_1,\dots,A_m) = \sum_{\{V_1,\dots,V_k\}} \sum_{n\geq m} \sumtwo{n_1\geq0, \dots,
n_k\geq0}{n_1+\dots+n_k \leq n-m} \prod_{\ell=1}^k \Bigl[ \frac{(|V_\ell|+n_\ell)!}{n_\ell!}
\int\dd\mu(A_1') \dots \int\dd\mu(A_{n_\ell}') \\
\varphi \bigl( (A_i)_{i\in V_\ell}, A_1', \dots,
A_{n_\ell}' \bigr) \Bigr] \, \frac1{p!} \int\dd\mu(A_1') \dots \int\dd\mu(A_p') \sum_{G\in\caG_p}
\prod_{(i,j) \in G} \zeta(A_i',A_j'),
\end{multline}
where $p = n - m - \sum_{i=1}^k n_i$. All terms converge absolutely and uniformly in $n$ because of
Theorems \ref{thmclexp} and \ref{thmdecay}; this allows to perform the sum
over $n$, and we obtain Theorem \ref{thmcorr}.
\end{proof}

\section{Illustrations}
\label{secill}

The first two examples are classical interacting particles and lattice polymers --- they are well-known
but nevertheless constitute nice illustrations of the results above. The last example deals with
quantum interacting particles and is more involved.

\subsection{Classical interacting gas}

We consider a gas of particles that are subject to pair interactions. Let $\bbA = \Lambda$ be a bounded
subset of $\bbR^d$ and $\dd\mu(x) = z \dd x$, where $z \in \bbR_+$ is the fugacity and $\dd x$ is the
Lebesgue measure. Let $\beta$ be the inverse temperature and $U(x-y) \geq 0$ represent the interactions between
particles at positions $x,y\in\Lambda$. Setting $\zeta(x,y) = \e{-\beta U(x-y)}-1$, the partition
function is given by \eqref{deffpart}. We choose $a(x)=1$ and we easily check that the
criterion \eqref{KPcrit1} holds whenever
\be
\label{condconv}
z \int\dd x (1-\e{-\beta U(x)}) \leq \e{-1}.
\ee
This condition for the convergence of the cluster expansion is well-known, see \cite{Rue} Chapter 4.
One then obtains an expression for the thermodynamic pressure $p(\beta,z)$, namely
\be
\beta \, p(\beta,z) = z + \sum_{n\geq1} z^{n+1} \int\dd x_1 \dots \int\dd x_n \varphi(0,x_1,\dots,x_n)
\ee
where integrals are over $\bbR^d$. This expression is absolutely convergent because of \eqref{b1}.
Furthermore the function $\varphi(\cdot)$ is analytic in $\beta$; by Vitali convergence theorem (see e.g.\
\cite{Sim} Theorem V.2.7 in the context of statistical physics), $p(\beta,z)$ is analytic in $\beta,z$
when the condition \eqref{condconv} is satisfied.

We study now the correlations. For bounded $\Lambda$, we consider the functions
\ba
&\rho_1(x_1) = \frac1Z \sum_{n\geq1} \frac{z^n}{(n-1)!} \int\dd x_2 \dots \int\dd x_n
\prod_{1\leq i<j\leq n} \bigl( 1 + \zeta(x_i,x_j) \bigr), \nn\\
&\rho_2(x_1,x_2) = \frac1Z \sum_{n\geq2} \frac{z^n}{(n-2)!} \int\dd x_3 \dots \int\dd x_n
\prod_{1\leq i<j\leq n} \bigl( 1 + \zeta(x_i,x_j) \bigr). \nn
\end{align}
By Theorem \ref{thmcorr}, the truncated two-point correlation function $\rho_2^{\text t}(x_1,x_2)$ is
given by
\ba
\rho_2^{\text t}(x_1,x_2) &= \rho_2(x_1,x_2) - \rho_1(x_1) \rho_1(x_2) \nn\\
&= \sum_{n\geq2} n(n-1) z^n \int\dd x_3 \dots \int\dd x_n \; \varphi(x_1,\dots,x_n).
\end{align}
We use Theorem \ref{thmdecay} (with $\gamma=1$) to get a bound for the decay of correlations. Let
$b\equiv0$ and $c(x)\geq0$ satisfy the triangle inequality, such that
\be
\label{condconv2}
z \int\dd x \bigl( 1 - \e{-\beta U(x)} \bigr) \e{c(x)} \leq \e{-1}.
\ee
We have
\ba
\e{c(x_1-x_2)} |\rho_2^{\text t}(x_1,x_2)| &\leq \sum_{n\geq2} n(n-1) z^n \int\dd x_3 \dots \int\dd x_n
|\varphi(x_1,\dots,x_n)| \e{c(x_1,\dots,x_n)} \nn\\
&\leq \e3 \bigl\{ 1 + \bigl( 1 - \e{-\beta U(x_1-x_2)} \bigr) \e{c(x_1-x_2)} \bigr\}.
\end{align}
The right side converges to $\e3$ as $|x_1-x_2| \to \infty$. This shows that
\be
|\rho_2^{\text t}(x_1,x_2)| \leq \const \e{-c(x_1-x_2)}
\ee
for all functions $c$ satisfying \eqref{condconv2}.

\subsection{Polymer models}

A polymer is a connected subset of $\bbZ^d$. Let $\bbA$ be the set of polymers in a finite set $\Lambda
\subset \bbZ^d$. The measure $\mu$ is taken to be the counting measure multiplied by a weight $w(A)$
satisfying $|w(A)| \leq \e{-\eta |A|}$ with $\eta = 2\log(2d\phi) + \phi^{-1}$. Here
$\phi=\frac{\sqrt5+1}2$ is the Golden Ratio. Polymers interact through a condition
of non-intersection, that is, $\zeta(A,A')$ is $-1$ if $A\cap A' \neq \emptyset$, and is 0 otherwise.

To check the criterion \eqref{KPcrit1}, we choose $a(A) = \phi^{-1} |A|$. It is enough to consider
the case where $A=\{0\}$. If $A$ is a connected set, there exists a closed walk with nearest-neighbor
jumps whose support is $A$, and whose length is at most $2|A|$. This can be seen by induction: knowing
the walk for $A$, it is easy to construct one for $A\cup\{x\}$. The number of connected sets of
cardinality $n$ that contain the origin is therefore smaller than the number of walks of length $2n$
starting at the origin, which is equal to $(2d)^{2n}$. The left side of \eqref{KPcrit1} is
bounded by $\sum_{n\geq1} (2d)^{2n} \e{-(\eta-\phi^{-1}) n}$ and this is equal to $\phi^{-1}$.

The cluster expansion provides absolutely convergent series for thermodynamic quantities.
Many physical models can be mapped onto a polymer model, and Theorem \ref{thmdecay} typically
provides informations on correlation functions.

\subsection{Quantum interacting gas}

The description of a gas of quantum particles in a bounded domain $\Lambda \subset \bbR^d$ should start
with the state space of the system, which is the Fock space $\caF_\pm = \bigoplus_{N\geq0} P_\pm(L^2(\Lambda))^{\otimes N}$, where $P_+$ (resp.\
$P_-$) is the projector onto symmetric (resp.\ antisymmetric) functions. Next one should introduce the Hamiltonian (Laplacian for the kinetic energy of the
particles, and another operator for the interactions). One could then write down the partition
function $Z = \Tr \e{-\beta (H-\mu N)}$.

It serves better our purpose to define the model in the Feynman-Kac representation. See \cite{Gin} for
a complete description of this representation, and for the definition of the Wiener measure, to be used below. Results of this section can actually be found in \cite{Gin},
but the cluster expansion used there is very intricate.

We start with the partition function
\bm
\label{fpartquant}
Z = \sum_{N\geq0} \frac{z^N}{N!} \int_\Lambda \dd x_1 \dots \int_\Lambda \dd x_N \sum_{\pi\in S_N}
\varepsilon^{|\pi|} \int\dd W_{x_1x_{\pi(1)}}^\beta(\omega_1) \dots \int\dd
W_{x_Nx_{\pi(N)}}^\beta(\omega_N) \\
\prod_{i=1}^N \upchi_\Lambda(\omega_i) \prod_{1\leq i<j\leq N}
\exp\Bigl\{ -\int_0^\beta U \bigl( \omega_i(t) - \omega_j(t) \bigr) \dd t \Bigr\}.
\end{multline}
Here, $S_N$ is the permutation group of $N$ elements, $\varepsilon=1$ for bosons and $\varepsilon=-1$
for fermions, $|\pi|$ is the number of transpositions of $\pi$, $\upchi_\Lambda(\omega)$ is 1 if
$\omega(t) \in \Lambda$ for all $0\leq t\leq\beta$ and is 0 otherwise, and $U(x)$ represents a
pair interaction potential. For simplicity we assume that $U$ is nonnegative and summable. Recall
that the Wiener measure $\dd W_{xy}^\beta(\omega)$ satisfies the following property; if
$0<t_1<\dots<t_n<\beta$ and if $f$ is a function $\bbR^{nd} \to \bbR$, we have
\bm
\int\dd W_{xy}^\beta(\omega) f \bigl( \omega(t_1), \dots, \omega(t_n) \bigr) \\
= \int_{\bbR^d} \dd x_1 \dots
\int_{\bbR^d} \dd x_n \psi_{t_1}(x_1-x) \psi_{t_2-t_1}(x_2-x_1) \dots \psi_{\beta-t_n}(y-x_n)
f(x_1,\dots,x_n),
\end{multline}
where $\psi_t(x)$ is the Gaussian
\be
\psi_t(x) = (2\pi t)^{-d/2} \e{-x^2/2t}.
\ee

The first step is to derive an expression for the partition function that is of the form
\eqref{deffpart}.
A permutation $\pi\in S_N$ can be decomposed in $k$ cycles of lengths $\ell_1,\dots\ell_k$ whose sum is
$N$. Given $\ell_1,\dots,\ell_k$, there are
$$
\frac1{k!} \frac{N!}{\prod_{i=1}^k \ell_i}
$$
corresponding permutations. Furthermore, the integrations over $x_1,\dots,x_\ell$ and over
$\omega_1$, \dots, $\omega_\ell$, where $\omega_j$ is a path from $x_j$ to $x_{j+1}$ (with $x_{\ell+1}
\equiv x_1$), can be performed by a single integration over $x$, followed by an integration over a path
$\omega$ with the measure $W_{xx}^{\ell\beta}$. We denote $\bsomega = (\ell,x,\omega)$ where $\ell$ is
a positive integer, $x\in\Lambda$, and $\omega$ is a path $[0,\ell\beta] \mapsto \bbR^d$ with
$\omega(0) = \omega(\ell\beta) = x$. We consider the following measure $\mu$
\bm
\dd\mu(\bsomega) = \frac{z^\ell \varepsilon^{\ell+1}}\ell \, \dd x \; \dd W_{xx}^{\ell\beta}(\omega)
\;\upchi_\Lambda(\omega) \\
\prod_{0\leq m<n\leq \ell-1} \exp\Bigl\{
-\int_0^\beta U \bigl( \omega(m\beta+t) - \omega(n\beta+t) \bigr) \dd t \Bigr\}.
\end{multline}
The partition function \eqref{fpartquant} can be written as
\bm
Z = \sum_{N\geq0} \frac1{N!} \int\dd\mu(\bsomega_1) \dots \int\dd\mu(\bsomega_N) \\
\prod_{1\leq i<j\leq
N} \exp\Bigl\{ -\sum_{m=0}^{\ell_i-1} \sum_{n=0}^{\ell_j-1} \int_0^\beta U \bigl( \omega_i(m\beta+t)
- \omega_j(n\beta+t) \bigr) \dd t \Bigr\}.
\end{multline}
We obtain \eqref{deffpart} by setting
\be
\zeta(\bsomega,\bsomega') = \exp\Bigl\{ -\sum_{m=0}^{\ell-1} \sum_{n=0}^{\ell'-1} \int_0^\beta U
\bigl( \omega(m\beta+t) - \omega'(n\beta+t) \bigr) \dd t \Bigr\} - 1.
\ee

We now establish the criterion \eqref{KPcrit1}. The condition below involves $z$, $\beta$, and $U$,
see \eqref{condconvquant}, and is not the most general that can be achieved. It is enough for the
purpose of an illustration of the use of cluster expansions, however.

We take $a(\bsomega)=(-\log z)\ell$. Since $1-\e{-t} \leq t$, we have
\bm
\int\dd|\mu|(\bsomega') |\zeta(\bsomega,\bsomega')| z^{-\ell'} \leq \sum_{\ell'\geq1}
\frac1{\ell'} \int\dd x' \int\dd W_{x'x'}^{\ell'\beta}(\omega') \\
\sum_{m=0}^{\ell-1}
\sum_{n=0}^{\ell'-1} \int_0^\beta U \bigl( \omega(m\beta+t) - \omega'(n\beta+t) \bigr) \dd t.
\end{multline}
Now for all $y$ and all $0<t'<\ell'\beta$, it is not hard to see that
\be
\int\dd x' \int\dd W_{x'x'}^{\ell'\beta}(\omega') U \bigl( y - \omega'(t') \bigr) =
(2\pi\ell'\beta)^{-d/2} \int U(x) \dd x.
\ee
We get then
\be
\int\dd|\mu|(\bsomega') |\zeta(\bsomega,\bsomega')| z^{-\ell'} \leq \frac{\ell\beta}{(2\pi\beta)^{d/2}} \int U(x) \dd
x \sum_{\ell'\geq1} {\ell'}^{-d/2}.
\ee
The criterion is fulfilled if the right side of the expression above is smaller than
$(-\log z)\ell$, that is,
if
\be
\label{condconvquant}
\frac\beta{(2\pi\beta)^{d/2}} \int U(x) \dd x \sum_{\ell\geq1} \ell^{-d/2} \leq -\log z.
\ee
This assumes in particular that $z \leq 1$. The thermodynamic pressure is analytic in $z,\beta$
in the range of parameters where \eqref{condconvquant} holds, and no condensation takes place, whether
classical or of the Bose-Einstein type.

Correlations are given by `reduced density matrices' and require more efforts.

\end{document}